\documentclass[twocolumn,trackchanges,resetfootnote]{aastex701}

\usepackage{CJK}
\usepackage{xcolor}
\usepackage{amsmath}
\usepackage{amssymb}
\usepackage{hyperref}

\definecolor{hl}{HTML}{000000}

\definecolor{rf}{HTML}{000000}
\newcommand{\rf}[1]{\textcolor{rf}{#1}}

\begin{document}
\begin{CJK}{UTF8}{gbsn}

\title{Planetary Accretion Is Less Frequent in Wide Binaries: Evidence from Metal-Enriched White Dwarfs in DESI DR1}

\author[0000-0002-8808-4282]{Siyi Xu (许\CJKfamily{bsmi}偲\CJKfamily{gbsn}艺)} 
\affil{Gemini Observatory/NSF NOIRLab, 950 N Cherry Ave, Tucson, AZ 85719, USA}
\email[show]{siyi.xu@noirlab.edu}

\author[0000-0002-3553-9474]{Laura. K. Rogers}
\affil{NSF NOIRLab, 950 N Cherry Ave, Tucson, AZ 85719, USA}
\email[]{laura.rogers@noirlab.edu}

\author[0000-0002-5758-150X]{Joan Najita}
\affil{NSF NOIRLab, 950 N Cherry Ave, Tucson, AZ 85719, USA}
\email[]{laura.rogers@noirlab.edu}

\author[0000-0002-2761-3005]{Boris T. G{\"a}nsicke}
\affil{Department of Physics, University of Warwick, Coventry CV4 7AL, UK}
\email[]{}

\author[0000-0002-0864-9432]{Paula Izquierdo}
\affil{Department of Physics, University of Warwick, Coventry CV4 7AL, UK}
\email[]{paula.izquierdo-sanchez@warwick.ac.uk}

\author[0000-0003-2644-135X]{Sergey E. Koposov}
\affil{Institute for Astronomy, University of Edinburgh, Royal Observatory, Blackford Hill, Edinburgh EH9 3HJ, UK}
\affiliation{Institute of Astronomy, University of Cambridge, Madingley Road,
Cambridge CB3 0HA, UK}
\email[]{skoposov@ed.ac.uk}

\author[0000-0003-1543-5405]{Christopher J. Manser}
\affil{Astrophysics Group, Department of Physics, Imperial College London, Prince Consort Rd, London SW7 2AZ, UK}
\email[]{}

\author[0000-0001-6515-9854]{Andrew Swan}
\affil{Department of Physics, University of Warwick, Coventry CV4 7AL, UK}
\email[]{andrew.swan@warwick.ac.uk}

\author[0000-0002-7393-3595]{Sandford	Nathan}
\affil{Department of Astronomy \& Astrophysics, University of Toronto, Toronto, ON M5S 3H4, Canada}
\email[]{nathan.sandford@utoronto.ca}

\author[0000-0002-5762-7571]{Wang	Wenting}
\affil{Department of Astronomy, School of Physics and Astronomy, Shanghai Jiao Tong University, Shanghai 200240, China}
\email[]{wenting.wang@sjtu.edu.cn}

\author[0000-0000-0000-0000,gname='Jessica Nicole', sname='Aguilar']{J.~Aguilar}
\affiliation{Lawrence Berkeley National Laboratory, 1 Cyclotron Road, Berkeley, CA 94720, USA}
\email{jaguilar@lbl.gov}

\author[0000-0001-6098-7247, gname='Steven', sname='Ahlen']{S.~Ahlen}
\affiliation{Department of Physics, Boston University, 590 Commonwealth Avenue, Boston, MA 02215 USA}
\email{ahlen@bu.edu}

\author[0000-0002-0084-572X, gname='Carlos ', sname='Allende Prieto']{C.~Allende~Prieto}
\affiliation{Departamento de Astrof\'{\i}sica, Universidad de La Laguna (ULL), E-38206, La Laguna, Tenerife, Spain}
\affiliation{Instituto de Astrof\'{\i}sica de Canarias, C/ V\'{\i}a L\'{a}ctea, s/n, E-38205 La Laguna, Tenerife, Spain}
\email{carlos.allende.prieto@iac.es}

\author[0000-0001-9712-0006, gname='Davide', sname='Bianchi']{D.~Bianchi}
\affiliation{Dipartimento di Fisica ``Aldo Pontremoli'', Universit\`a degli Studi di Milano, Via Celoria 16, I-20133 Milano, Italy}
\affiliation{INAF-Osservatorio Astronomico di Brera, Via Brera 28, 20122 Milano, Italy}
\email{davide.bianchi1@unimi.it}

\author[0000-0000-0000-0000,gname='David', sname='Brooks']{D.~Brooks}
\affiliation{Department of Physics \& Astronomy, University College London, Gower Street, London, WC1E 6BT, UK}
\email{david.brooks@ucl.ac.uk}

\author[0000-0000-0000-0000,gname='Todd', sname='Claybaugh']{T.~Claybaugh}
\affiliation{Lawrence Berkeley National Laboratory, 1 Cyclotron Road, Berkeley, CA 94720, USA}
\email{tmclaybaugh@lbl.gov}

\author[0000-0002-1769-1640, gname='Axel ', sname='de la Macorra']{A.~de la Macorra}
\affiliation{Instituto de F\'{\i}sica, Universidad Nacional Aut\'{o}noma de M\'{e}xico,  Circuito de la Investigaci\'{o}n Cient\'{\i}fica, Ciudad Universitaria, Cd. de M\'{e}xico  C.~P.~04510,  M\'{e}xico}
\email{macorra@fisica.unam.mx}

\author[0000-0002-2890-3725, gname='Jaime E.', sname='Forero-Romero']{J.~E.~Forero-Romero}
\affiliation{Departamento de F\'isica, Universidad de los Andes, Cra. 1 No. 18A-10, Edificio Ip, CP 111711, Bogot\'a, Colombia}
\affiliation{Observatorio Astron\'omico, Universidad de los Andes, Cra. 1 No. 18A-10, Edificio H, CP 111711 Bogot\'a, Colombia}
\email{je.forero@uniandes.edu.co}

\author[0000-0003-3142-233X, gname='Satya ', sname='Gontcho A Gontcho']{Satya~{Gontcho A Gontcho}}
\affiliation{University of Virginia, Department of Astronomy, Charlottesville, VA 22904, USA}
\email{satya@virginia.edu}

\author[0000-0000-0000-0000,gname='Gaston', sname='Gutierrez']{G.~Gutierrez}
\affiliation{Fermi National Accelerator Laboratory, PO Box 500, Batavia, IL 60510, USA}
\email{gaston@fnal.gov}

\author[0000-0003-0201-5241, gname='Dick', sname='Joyce']{R.~Joyce}
\affiliation{NSF NOIRLab, 950 N. Cherry Ave., Tucson, AZ 85719, USA}
\email{richard.joyce@noirlab.edu}

\author[0000-0003-1838-8528, gname='Martin', sname='Landriau']{M.~Landriau}
\affiliation{Lawrence Berkeley National Laboratory, 1 Cyclotron Road, Berkeley, CA 94720, USA}
\email{mlandriau@lbl.gov}

\author[0000-0001-7178-8868, gname='Laurent', sname='Le Guillou']{L.~Le~Guillou}
\affiliation{Sorbonne Universit\'{e}, CNRS/IN2P3, Laboratoire de Physique Nucl\'{e}aire et de Hautes Energies (LPNHE), FR-75005 Paris, France}
\email{llg@lpnhe.in2p3.fr}

\author[0000-0002-1125-7384, gname='Aaron', sname='Meisner']{A.~Meisner}
\affiliation{NSF NOIRLab, 950 N. Cherry Ave., Tucson, AZ 85719, USA}
\email{aaron.meisner@noirlab.edu}

\author[0000-0000-0000-0000,gname='Ramon', sname='Miquel']{R.~Miquel}
\affiliation{Instituci\'{o} Catalana de Recerca i Estudis Avan\c{c}ats, Passeig de Llu\'{\i}s Companys, 23, 08010 Barcelona, Spain}
\affiliation{Institut de F\'{i}sica d’Altes Energies (IFAE), The Barcelona Institute of Science and Technology, Edifici Cn, Campus UAB, 08193, Bellaterra (Barcelona), Spain}
\email{rmiquel@ifae.es}

\author[0000-0002-0644-5727, gname='Will', sname='Percival']{W.~J.~Percival}
\affiliation{Department of Physics and Astronomy, University of Waterloo, 200 University Ave W, Waterloo, ON N2L 3G1, Canada}
\affiliation{Perimeter Institute for Theoretical Physics, 31 Caroline St. North, Waterloo, ON N2L 2Y5, Canada}
\affiliation{Waterloo Centre for Astrophysics, University of Waterloo, 200 University Ave W, Waterloo, ON N2L 3G1, Canada}
\email{will.percival@uwaterloo.ca}

\author[0000-0001-7145-8674, gname='Francisco', sname='Prada']{F.~Prada}
\affiliation{Instituto de Astrof\'{i}sica de Andaluc\'{i}a (CSIC), Glorieta de la Astronom\'{i}a, s/n, E-18008 Granada, Spain}
\email{fprada@iaa.es}

\author[0000-0001-6979-0125, gname='Ignasi', sname='Pérez-Ràfols']{I.~P\'erez-R\`afols}
\affiliation{Departament de F\'isica, EEBE, Universitat Polit\`ecnica de Catalunya, c/Eduard Maristany 10, 08930 Barcelona, Spain}
\email{ignasi.perez.rafols@upc.edu}

\author[0000-0000-0000-0000,gname='Graziano', sname='Rossi']{G.~Rossi}
\affiliation{Department of Physics and Astronomy, Sejong University, 209 Neungdong-ro, Gwangjin-gu, Seoul 05006, Republic of Korea}
\email{graziano@sejong.ac.kr}

\author[0000-0002-9646-8198, gname='Eusebio', sname='Sanchez']{E.~Sanchez}
\affiliation{CIEMAT, Avenida Complutense 40, E-28040 Madrid, Spain}
\email{eusebio.sanchez@ciemat.es}

\author[0000-0000-0000-0000,gname='David', sname='Schlegel']{D.~Schlegel}
\affiliation{Lawrence Berkeley National Laboratory, 1 Cyclotron Road, Berkeley, CA 94720, USA}
\email{djschlegel@lbl.gov}

\author[0000-0002-3461-0320, gname='Joseph Harry', sname='Silber']{J.~Silber}
\affiliation{Lawrence Berkeley National Laboratory, 1 Cyclotron Road, Berkeley, CA 94720, USA}
\email{jhsilber@lbl.gov}

\author[0000-0003-1704-0781, gname='Gregory', sname='Tarlé']{G.~Tarl\'{e}}
\affiliation{University of Michigan, 500 S. State Street, Ann Arbor, MI 48109, USA}
\email{gtarle@umich.edu}

\author[0000-0000-0000-0000,gname='Benjamin Alan', sname='Weaver']{B.~A.~Weaver}
\affiliation{NSF NOIRLab, 950 N. Cherry Ave., Tucson, AZ 85719, USA}
\email{benjamin.weaver@noirlab.edu}

\author[0000-0001-5381-4372, gname='Rongpu', sname='Zhou']{R.~Zhou}
\affiliation{Lawrence Berkeley National Laboratory, 1 Cyclotron Road, Berkeley, CA 94720, USA}
\email{rongpuzhou@lbl.gov}

\collaboration{all}{The DESI collaboration}

\begin{abstract}

Binary stars are common in the Galaxy, and understanding how stellar binarity influences the formation and evolution of planetary systems is an active area of research. In this study, we use metal-enriched white dwarfs in wide binaries as tracers of long-lived planetary systems. With Data Release~1 from the Dark Energy Spectroscopic Instrument (DESI), we find that the fraction of cool metal-enriched white dwarfs in wide binaries is \rf{9.8}\,$\pm$\,2.1\%, significantly lower (4.7\,$\sigma$) than the 20.5\,$\pm$\,0.9\% in a control sample of single systems. {\color{hl} Furthermore, we identify a tentative dependence of metal enrichment on projected separation and white dwarf effective temperature, where enrichment fraction decreases at smaller separations and lower temperatures.} \rf{These findings indicate that, compared to single stars, binary systems either start with smaller initial planetary reservoirs due to suppressed planetesimal formation or undergo more rapid depletion of planetary material during the initial part of the white dwarf stage.}
\end{abstract}

\keywords{\uat{Surveys}{1671} --- \uat{Exoplanet evolution}{491} --- \uat{Planetesimals}{159} -- \uat{Binary stars}{154} --- \uat{White dwarf stars}{1799}}


\section{Introduction} \label{sec: introduction}

Binary stars are a common occurrence. The binary fraction of solar-type stars is approximately 50\%, and this significantly increases for higher-mass stars \citep{Fuhrmann2017, Offner2023}. How does stellar multiplicity affect the architecture of planetary systems? During the planet formation process, {\color{hl} a companion can truncate the protoplanetary disk and limit the available disk mass, thereby inhibiting giant planet formation \citep{Harris2012}. Moreover, exoplanet demographic studies find that while planets are less common around close binaries, the planet occurrence rate becomes comparable to single stars for binaries with a separation larger than $\approx$ 200\,AU \citep[e.g.][]{Wang2014, MoeKratter2021, Clark2024}.}

In this study, we use \rf{cool} metal-enriched white dwarfs -- those that display elements heavier than He in their atmospheres \citep[e.g.][]{JuraYoung2014, Williams2024, Rogers2024a} -- to probe long-lived planetary systems. {\color{hl}The detection of metal enrichment requires recent accretion of planetary material, which in turn implies the presence of a planetary system -- specifically a reservoir of planetesimals -- around the white dwarf} \citep[e.g.][]{FrewenHansen2014, Veras2024, Pham2024}. {\color{hl} The detectability of metals in a given system depends on its abundance in the white dwarf atmosphere, the sensitivity of the instrument, and the atmospheric parameters of the white dwarf, especially the effective temperature and the dominant atmospheric composition.} 

\rf{Metal enrichment is frequently detected in single white dwarfs.} Using the High Resolution Echelle Spectrometer (HIRES) on the Keck Telescope, the occurrence rate of metal enrichment is found to be around 25\% in single systems with both hydrogen-dominated and helium-dominated atmospheres \citep{Zuckerman2003, Zuckerman2010}. \citet{Manser2024a} extended this work to cool (white dwarf effective temperature, T$_\mathrm{eff}$, $<11,500$\,K) helium-dominated white dwarfs and found a similar fraction of 21\,$\pm$\,3\% for metal enrichment using the white dwarf catalog from the DESI Early Data Release (EDR). This high level of enrichment appears to be sustained for a long time, with the mass accretion rate decreasing by no more than a factor of 10 over a white dwarf cooling age of between 1 and 8~Gyr (4,000\,K\,$\lesssim$\,T$_\mathrm{eff}$\,$\lesssim$\,8,000\,K, \citealt{BlouinXu2022}). 



Theoretical studies show that binaries can enhance the delivery of planetary material onto white dwarfs, via {\color{hl} the} \rf{von Zeipel-Kozai-Lidov} mechanism or perturbations from the Galactic tides \citep[e.g.][]{Bonsor2015, PetrovichMunoz2017, Stephan2017, Trierweiler2026}. Observationally, for white dwarfs in close binary systems ($<$ few AU), metal enrichment can occur due to {\color{hl} the accretion of stellar wind from a companion}. However, for more widely separated binaries, any metal enrichment must have a planetary origin \citep{Veras2018}. The closest metal enriched white dwarf Procyon b is in fact in a wide binary system and very likely hosts a planetary system \citep{Provencal2002}. Metal enrichment in wide binaries has been investigated in a few previous studies, but the overall sample size is small, on the order of a few dozen \citep{Zuckerman2014, Wilson2019, Williams2024, Noor2024}. 


In this study, we use a larger dataset from DESI, which is a robotic, fiber-fed, highly multiplexed spectroscopic survey with the Mayall 4-meter telescope at the Kitt Peak National Observatory \citep{DESI_2022}. DESI can simultaneously obtain spectra of 5,000 targets over a $\sim$ 3$^\circ$ field of view \citep{DESI_2016, Miller2024, Poppett2024}. Its primary goal is to determine the nature of dark energy, and it will obtain spectra for 63 million galaxies and quasars over 17,000\,sq.~deg. of the sky over an eight-year period \citep{Schlafly2023, Guy2023}. DESI has issued two public data releases: the EDR, covering the first six months of observations from commissioning and survey validation, and Data Release 1 (DR1; \citealt{DESI_DR1}), which includes data from the first 13 months of the main survey. Early DESI results have already found evidence for evolving dark energy \citep{DESI_2024_VII, DESI_2025}. The DR1 stellar catalog contains spectra, stellar parameter and radial velocity (RV) measurements of 4 million stars \citep{Koposov2026}. In this study, we use the DR1 white dwarf catalog \citep{Swan2026}, which contains \rf{44,409} white dwarfs and is the largest and most uniform spectroscopic catalog of white dwarfs.

The paper is organized as follows. The sample selection, including the wide binary sample and a control sample of single stars, is reported in Section~\ref{sec: sample}. The metal enrichment rate calculation and the characteristics of different samples are presented in Section~\ref{sec: results}. The discussion and conclusions are given in Sections~\ref{sec: discussion} and \ref{sec: conclusions}, respectively.

\section{Sample Selection \label{sec: sample}}

Compared to the 2,706 spectroscopically confirmed DESI EDR white dwarfs \citep{Manser2024b}, the DESI DR1 white dwarf catalog, with \rf{44,409} white dwarfs, represents a significant increase in white dwarf numbers \citep{Swan2026}. White dwarfs are observed primarily as science targets under the Milky Way Survey programs \citep{Cooper2023}, also as flux standards in the validation surveys \citep{Myers2023}, and in backup programs \citep{Dey2025}. A detailed description and characterization of the white dwarf catalog is presented in \citet{Swan2026}. The analysis presented in this work is only possible thanks to the large and homogeneous white dwarf sample provided by DESI.

Here, we focus on cool (T$_\mathrm{eff}$\,$<$\,11,500\,K) helium-dominated white dwarfs. Even though metal enrichment occurs over a range of white dwarf temperatures, the metal lines tend to be strongest in cool helium-atmosphere white dwarfs \citep[e.g.][]{Xu2024b}, {\color{hl}which increases the detectability of these features using low-resolution data. The pilot study for metal enrichment using EDR data has shown that DESI is a powerful survey for studying remnant planetary systems around white dwarfs \citep{Manser2024a}.}

Starting with the DESI DR1 white dwarf catalog from \citet{Swan2026}, we applied five additional selection criteria to ensure that the data quality is uniform and the spectral classification is reliable. We adopt the same column names as those in the DESI DR1 white dwarf catalog and follow their recommendations. (1) The probability of being a white dwarf from DESI is taken as $\mathtt{pwd\_DESI}$\,$\geq$\,0.5. (2) The confidence in the spectral type classification is taken as $\mathtt{specType\_confidence}$\,$\geq$\,0.8. (3) Expected fractional fiber contamination from field stars based on \textit{Gaia} data is taken as $\mathtt{fluxContamination\_DESI}$\,$\leq$\,0.001 -- this requires that at least 99.9\% of the light entering into the DESI fiber comes from the white dwarf. (4) The signal-to-noise ratio measured statistically in the blue arm is taken as $\mathtt{SNmeasured\_B}$\,$\geq$\,4\footnote{\rf{$\mathtt{SNmeasured\_B}$ represents the scatter per pixel, calculated empirically across 4550--4650~{\AA}. We consider this a more realistic representation of the data quality than the nominal $\mathtt{SN\_B}$ value \citep{Swan2026}.}} -- this ensures that there is sufficient signal in the blue arm, the region of the strongest metal lines for \rf{cool} metal-enriched white dwarfs, Ca H \& K. \rf{A threshold of 4 balances sample size against spectral classification reliability, as human classification becomes significantly discrepant below an $\mathtt{SNmeasured\_B}$ of 4.} (5) The white dwarf was observed in the dark or bright programs of the main survey, requiring $\mathtt{primary}$\,=\,True. Effectively, this excludes white dwarfs observed by DESI under other programs, which were targeted under different or serendipitous selection criteria (see more discussion about the primary sample in \citealt{Swan2026}). These five criteria reduce the DR1 white dwarf sample from \rf{44,409} to \rf{32,456}. 

We follow the procedures outlined in \citet{Manser2024a} to select cool helium-dominated white dwarfs (5,000\,$\lesssim$\,T$_\mathrm{eff}$\,$\lesssim$\,11,500\,K, with cooling ages, $\tau$, from approximately 460\,Myr\,$<$\,$\tau$\,$<$\,6.4\,Gyr from \citealt{Bedard2020}). These objects have the spectral types of DCs (which show no absorption features), and DQs (which show carbon dredge up features), while their metal-enriched counterparts are DZs (which show absorption lines from metals), DQZs (with carbon features being the strongest, followed by absorption from metals), and DZQs (with absorption from metals being the strongest feature, followed by carbon features). An additional \textit{Gaia} color cut 0\,$<$\,(G$_{BP}$\,$-$\,G$_{RP}$)\,$<$\,1 is applied to exclude contaminants, such as high-mass DQs (because they are likely merger products \citealt{Kawka2023}) and hydrogen-dominated DCs (because they do not have a helium-dominated atmosphere \citealt{Caron2023}). There are a total of \rf{2,805} cool helium-dominated white dwarfs that meet these criteria in DESI DR1.

\subsection{Wide Binary Sample \label{sec: sample_binary}}

To select white dwarfs in wide binaries, we cross-match the \rf{2,805} cool helium-dominated white dwarfs from DESI DR1 with the \textit{Gaia}~eDR3 wide binary catalog from \citet{El-Badry2021}, which uses the parallaxes and proper motions to produce a sample of spatially resolved binaries with projected separation from a few AU up to 1~pc. The main false positive in the binary catalog comes from chance alignments of background objects; therefore, we apply a stringent constraint to only include systems with a small probability of chance alignment, i.e., $\mathtt{R\_chance\_align}$\,$<$\,0.01. We also exclude white dwarfs with a \textit{Gaia} $\mathtt{RUWE}$ number higher than 1.25, which are likely to have a close companion \citep{Penoyre2022}, in addition to the wide companion already identified in \citet{El-Badry2021}. As a result, we have identified a total of \rf{215} systems, including \rf{183} in white dwarf--main sequence binaries, 30 white dwarf--white dwarf binaries, and two additional white dwarfs with unknown companion types\footnote{The unknown companions have `WD??' in the binary star classification, meaning that the companion has no \textit{Gaia} color information \citep{El-Badry2021}.}. See Table~\ref{tab: numbers} for a detailed breakdown of the white dwarf types. {\color{hl} Among the companions to these \rf{215} white dwarfs, 33 have available DESI DR1 spectra \citep{Koposov2024}}. The sample of metal-enriched white dwarfs in wide binaries is provided in Table~\ref{tab: WDMS}.

\begin{deluxetable*}{lcccccccccc}
\tablecaption{ \label{tab: numbers} Number of cool helium-dominated white dwarfs with different spectral types in wide binaries and single systems from DESI DR1. The calculation of metal enrichment fraction is discussed in Section~\ref{sec: enrichment frequency}.}
\tablehead{
\colhead{} &  \colhead{Wide Binary } & \colhead{Single System}}
\startdata
n$_\mathrm{DC}$ & \rf{139}& \rf{1661}\\
n$_\mathrm{DQ}$ & \rf{55}& \rf{387}\\
n$_\mathrm{DZ}$ & 21 & \rf{528}\\
n$_\mathrm{DQZ}$ & 0& 1\\
n$_\mathrm{DZQ }$& 0& 1\\
\\
Enrichment Fraction $f$ & \rf{9.8}$\,\pm$\,2.1\%& 20.6\,$\pm$\,0.9\% \\
  \enddata
\end{deluxetable*}

To further validate the wide binary selection criteria from \citet{El-Badry2021}, we compared the RVs of the binary components using results from \citet{Swan2026} and the literature, and they should match if they are truly associated. We identified \rf{482} white dwarfs in DESI DR1 with RV \rf{uncertainties $<$ 50\,km/s} that also have measured companion RVs from the literature. \rf{In comparison, the median uncertainty is 45\,km/s for the entire catalog \citep{Swan2026}.} These companion data were compiled from DESI DR1 \citep{Koposov2026}, SDSS \citep{APOGEE17, SDSSV, SDSS-DR14, SDSS-SSPP}, LAMOST \citep{LAMOST-LRS, LAMOST-MRS}, and {\it Gaia} \citep{GaiaDR3-RVS}. \rf{Although the individual RV uncertainties are large, our sample of 482 systems yields a strong correlation between the white dwarf and companion RVs, with a Pearson correlation coefficient of 0.677. This coefficient drops to 0.195 when randomly pairing the white dwarf with a nearby star. In addition, the difference between the RVs of the white dwarf and the companion peaks at 30\,km/s, consistent with the gravitational redshift of a typical white dwarf.} \rf{These tests reinforce} our confidence in the robustness of the wide binary sample.


\subsection{A Comparison Single White Dwarf Sample \label{sec: sample_single}}

We also compiled a control sample of cool helium-dominated white dwarfs in single systems from DESI DR1. Starting again with the \rf{2,805} systems in Section~\ref{sec: sample}, we excluded the \rf{215} wide binary systems identified in Section~\ref{sec: sample_binary} and the \rf{12} systems with a \textit{Gaia} $\mathtt{RUWE}$ number higher than 1.25, returning a total of \rf{2,578} high-confidence single white dwarfs. A breakdown of the number of white dwarfs in each spectral type is listed in Table~\ref{tab: numbers}. {\color{hl} We have performed additional checks and found that the single white dwarf and the wide binary samples are statistically indistinguishable across several key parameters -- including spectral types, the signal-to-noise ratio of the DESI spectra measured in the blue arm ($\mathtt{SNmeasured\_B}$), and the white dwarf effective temperature -- which further establishes the single-star sample as a suitable control. }

\section{Results \& Analysis \label{sec: results}}

\subsection{Metal Enrichment Fraction \label{sec: enrichment frequency}}

To calculate the metal enrichment fraction $f$, we can apply the equation following \citet{Manser2024a}:

$$ f = \frac{n_\mathrm{DZ} + n_\mathrm{DZQ} + n_\mathrm{DQZ}}{n_\mathrm{DC}+n_\mathrm{DQ}+n_\mathrm{DZ} + n_\mathrm{DZQ} + n_\mathrm{DQZ}}$$
Using the numbers in Table~\ref{tab: numbers}, we find that the fraction of metal enrichment {\color{hl} for white dwarfs in wide binaries and single systems are \rf{9.8}\,$\pm$\,2.1\% and 20.5\,$\pm$\,0.9\%, respectively, assuming Poisson statistics}. Therefore, the enrichment fraction in wide binaries differs from that in single systems by 4.7\,$\sigma$. If the wide binary sample selection is relaxed slightly, increasing the chance alignment parameter $\mathtt{R\_chance\_align}$ from 0.01 to 0.1, the metal enrichment rate becomes 10.0\,$\pm$\,2.1\% in wide binaries, still 4.6\,$\sigma$ lower than that of single white dwarfs. Regardless of the exact cut on the chance alignment parameter, the result remains robust: the occurrence of metal enrichment is significantly lower in wide binaries compared to single systems. 

Our final binary sample includes 30 white dwarf--white dwarf pairs\footnote{For all 30 systems, only one component of each white dwarf--white dwarf pair has an available DESI DR1 spectrum.}, two of which are enriched with metals; this makes the fraction of metal enrichment in white dwarf--white dwarf binaries 6.6\,$\pm$\,4.7\% (2/30), which is slightly lower than the 10.3\,$\pm$\,2.4\% (19/185) derived from white dwarf--main sequence binaries. A larger sample is needed to confirm whether the difference in enrichment fraction for white dwarf--white dwarf binaries compared to white dwarf--main sequence binaries is statistically significant. It is also worth noting that our single-system enrichment fraction of 20.5\,$\pm$\,0.9\% is consistent with the value of 21\,$\pm$\,3\%\footnote{\rf{The DESI EDR dataset includes both wide binaries and single systems. If we exclude the binaries identified by \citet{El-Badry2021}, the metal enrichment fraction for single systems becomes $48/219 = 21.9 \pm 3.2\%$, which remains consistent with the fraction reported here for DR1.}} reported in the DESI EDR analysis \citep{Manser2024a}. It shows that with the quality cuts presented in Section~\ref{sec: sample}, it is possible to extract a high-quality sample of white dwarfs from DESI DR1, even though the EDR data on average have longer exposure times and therefore, higher signal-to-noise ratios  \citep{Swan2026}.

\subsection{\rf{SDSS Comparison Sample}}

\rf{The Sloan Digital Sky Survey (SDSS) has observed a large number of white dwarfs \citep{Ahumuda2020,SDSSV}. \citet{Lizana-Vidal2026} compiled a homogeneous sample of hydrogen-dominated white dwarfs from SDSS and found a metal enrichment fraction below $\sim$1\% for both single systems and wide binaries. The low enrichment fraction is likely due to the metal lines being a lot weaker in hydrogen-dominated white dwarfs compared to their helium counterparts -- the focus of this study. As an additional exercise comparing cool helium-dominated white dwarfs in SDSS and DESI, we start with the Gaia-SDSS DR16 sample compiled by \citet{Gentile-Fusillo2021b} and applied the same set of criteria listed in Sections~\ref{sec: sample_binary} and \ref{sec: sample_single} to wide binaries and single systems, respectively. We calculate the corresponding metal enrichment fractions, $f_\mathrm{SDSS, single}$ and $f_\mathrm{SDSS, binary}$, as}

\rf{
\begin{equation*}
f_\mathrm{SDSS, single} = \frac{839}{3665} = 22.9\pm0.8\%
\end{equation*}
\begin{equation*}
f_\mathrm{SDSS, binary} = \frac{13}{143} = 9.1\pm2.5\%
\end{equation*}
}
\rf{The SDSS sample also shows that the fraction of metal enrichment in cool, helium-dominated white dwarfs is lower in binaries compared to single stars, further validating the DESI results. In addition, for single systems, the metal enrichment fraction is consistent between DESI and SDSS, further demonstrating the reliability of measuring metal enrichment in helium-dominated white dwarfs using low-resolution data.}

\subsection{Comparison of White Dwarf Properties \label{sec: WD properties}}

\begin{figure}
\plotone{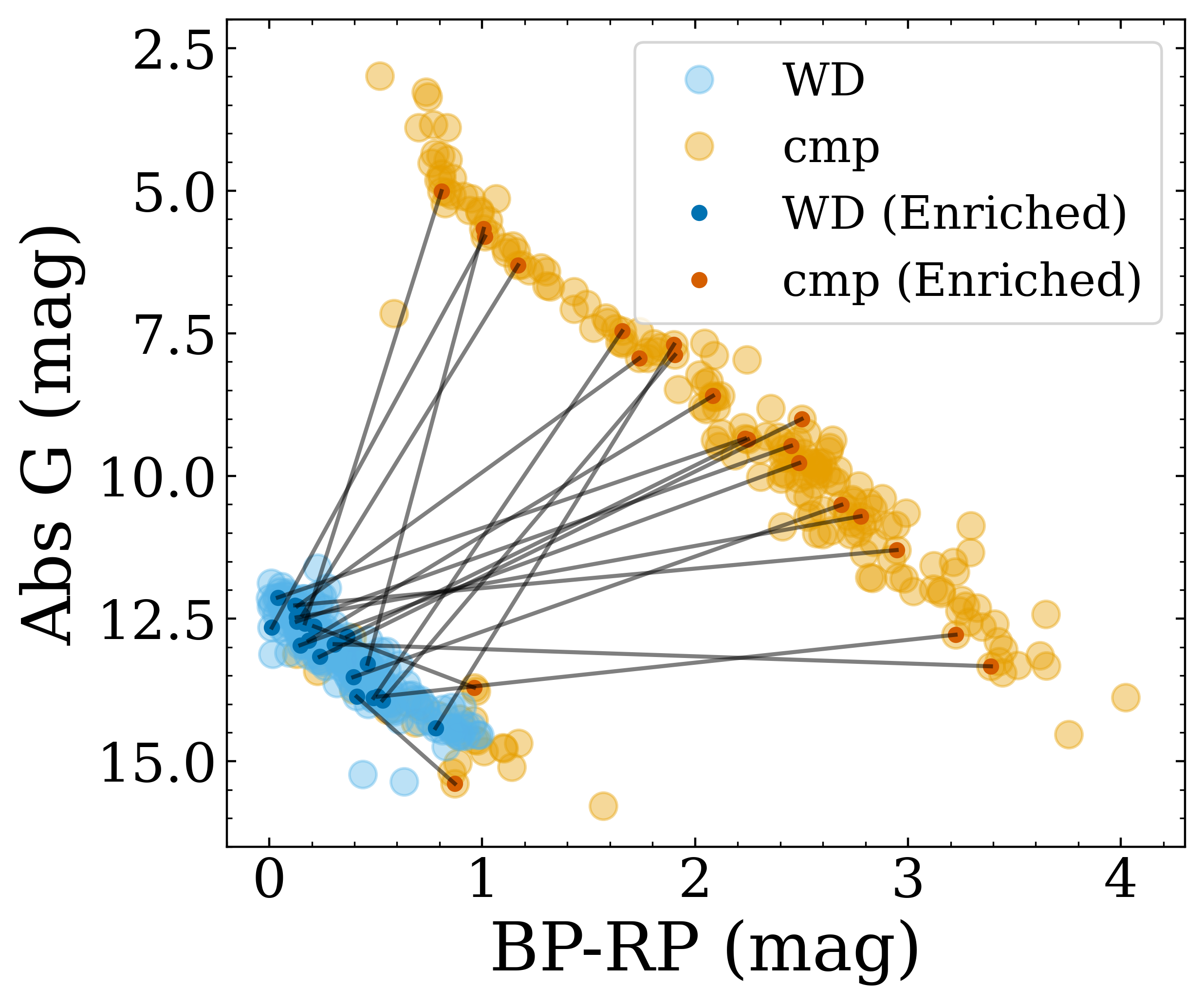}
\caption{\textit{Gaia} HR diagram of cool helium-dominated white dwarfs (WD) and their binary companion (cmp). For the white dwarf--white dwarf binaries, we color code the one with a DESI spectrum as light blue, and the other without a DESI spectrum as orange. The metal-enriched systems are shown in darker colors and are connected to their companions by lines. 
}
\label{fig: HR}
\end{figure}

In this section, we examine properties of the white dwarf samples studied here. Starting with the binary sample, their positions in the \textit{Gaia} Hertzsprung-Russell (HR) diagram are shown in Figure~\ref{fig: HR}. The enriched and non-enriched white dwarfs occupy a similar region in the HR diagram. 

\begin{figure*}
\plottwo{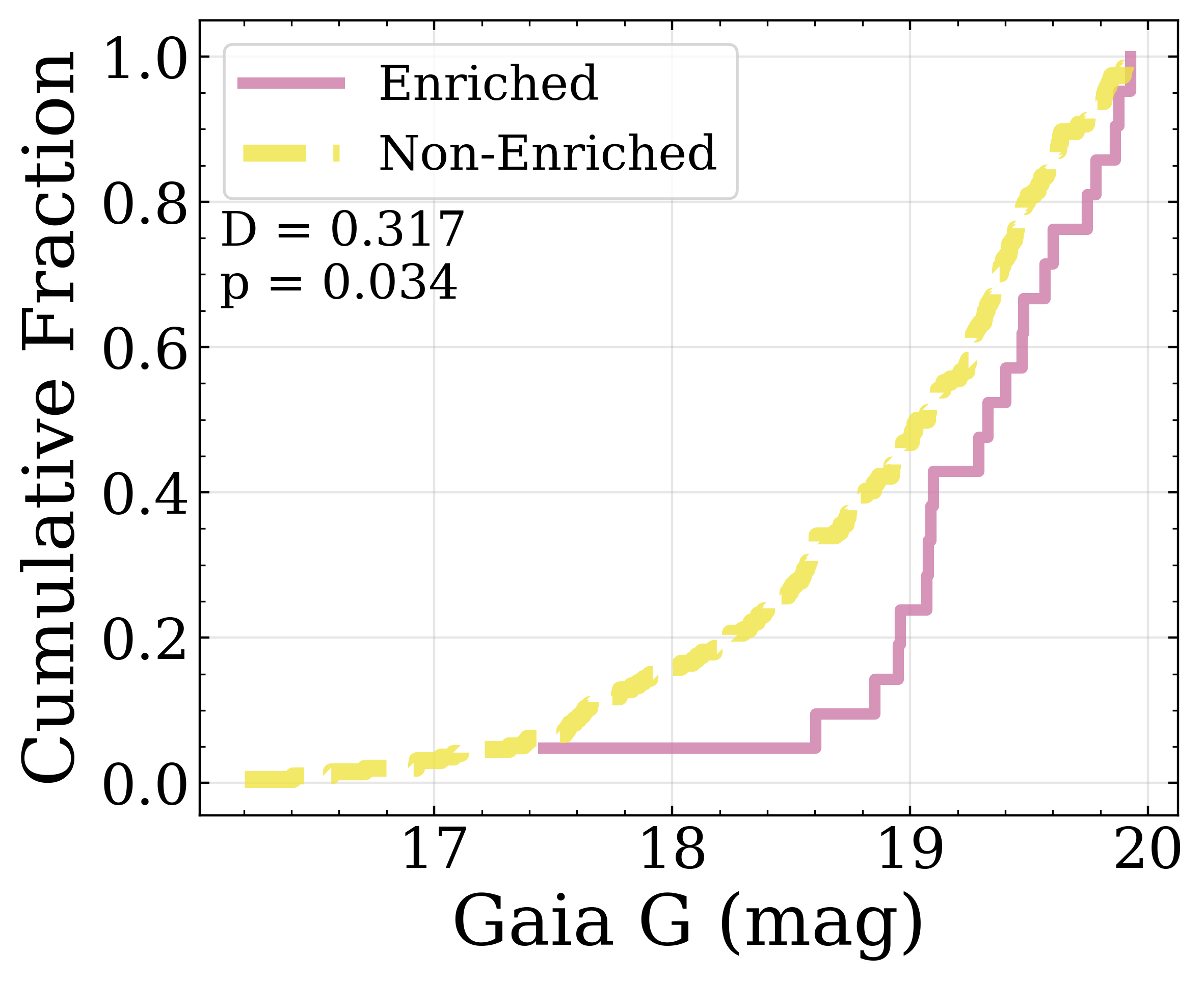}{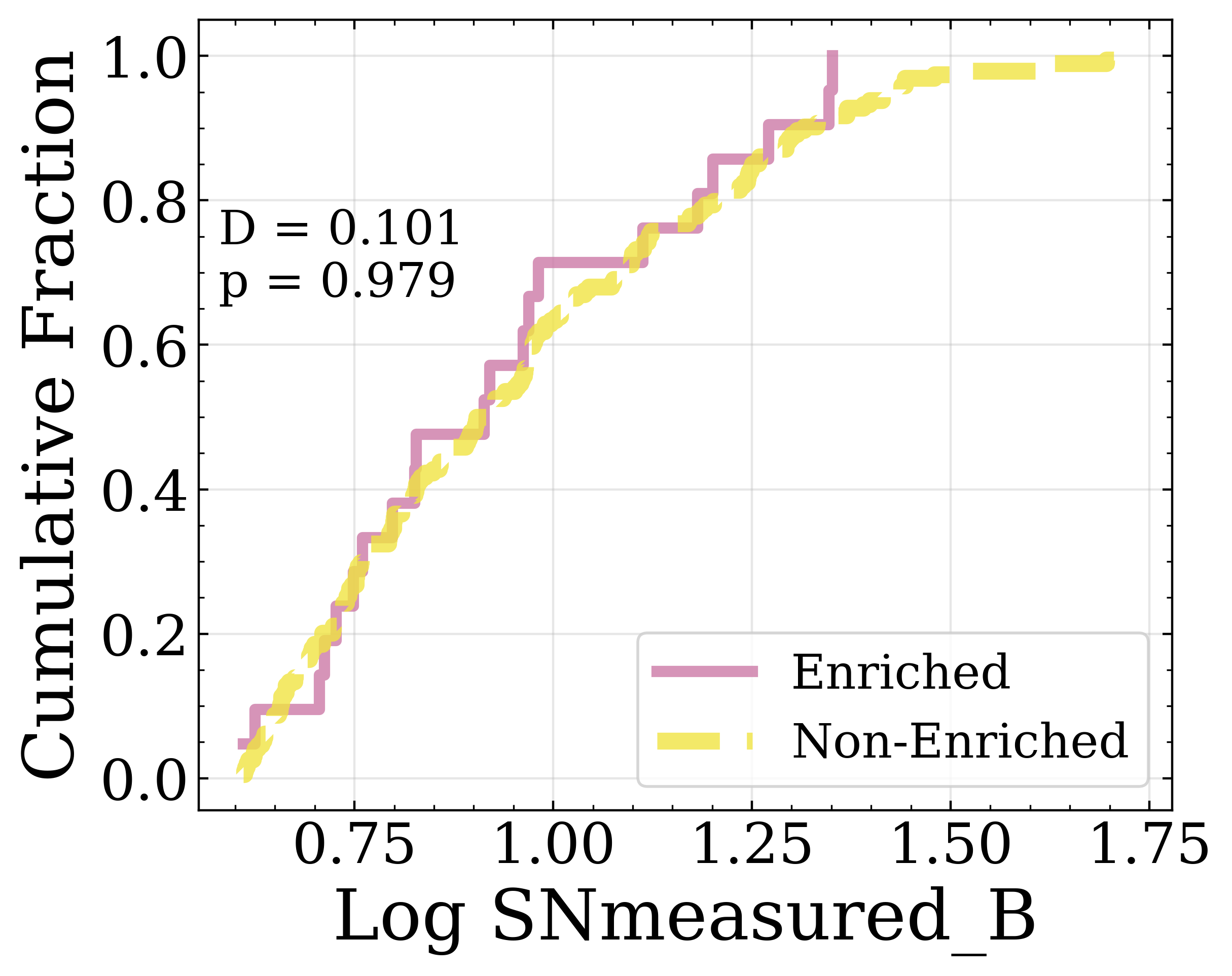}
\plottwo{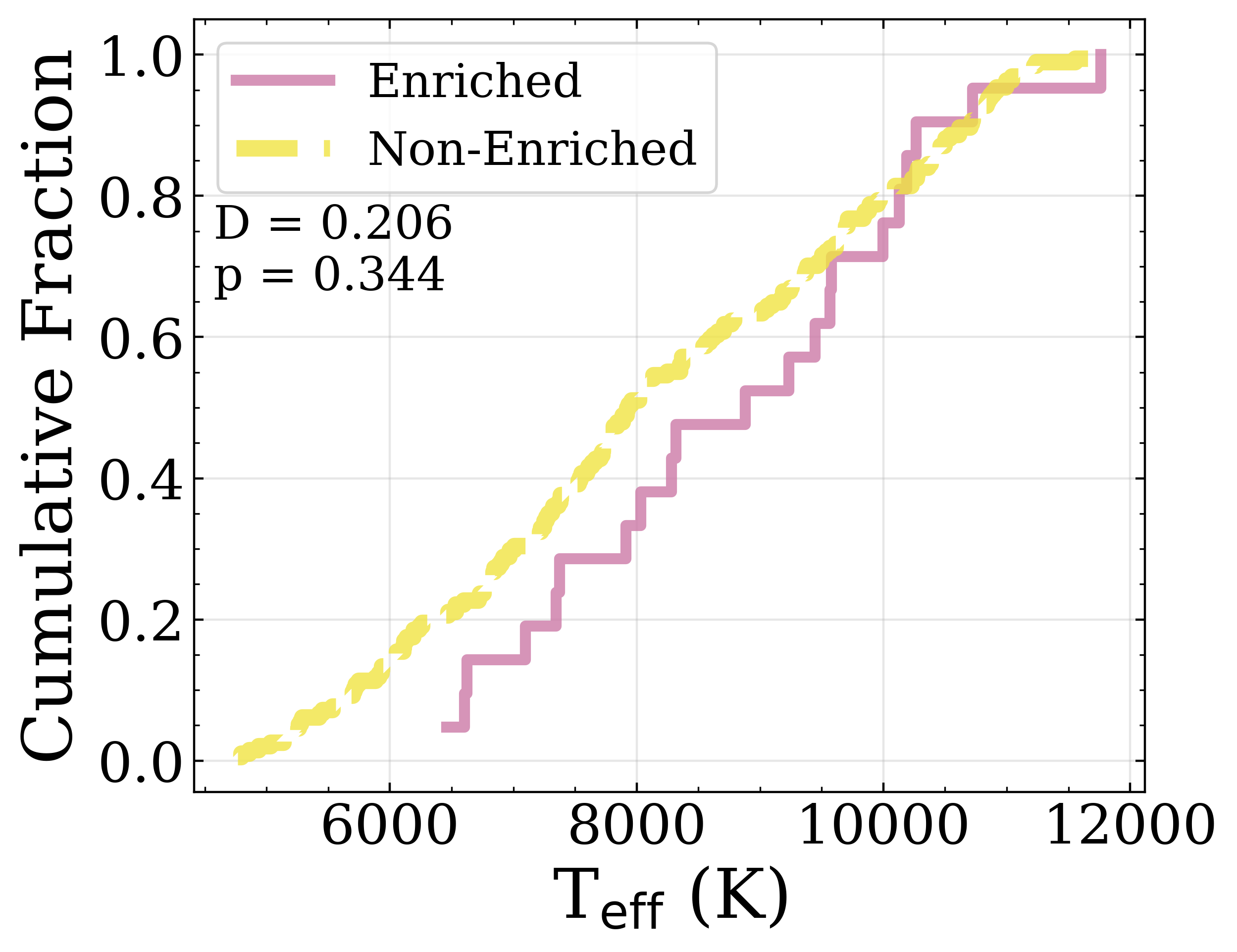}{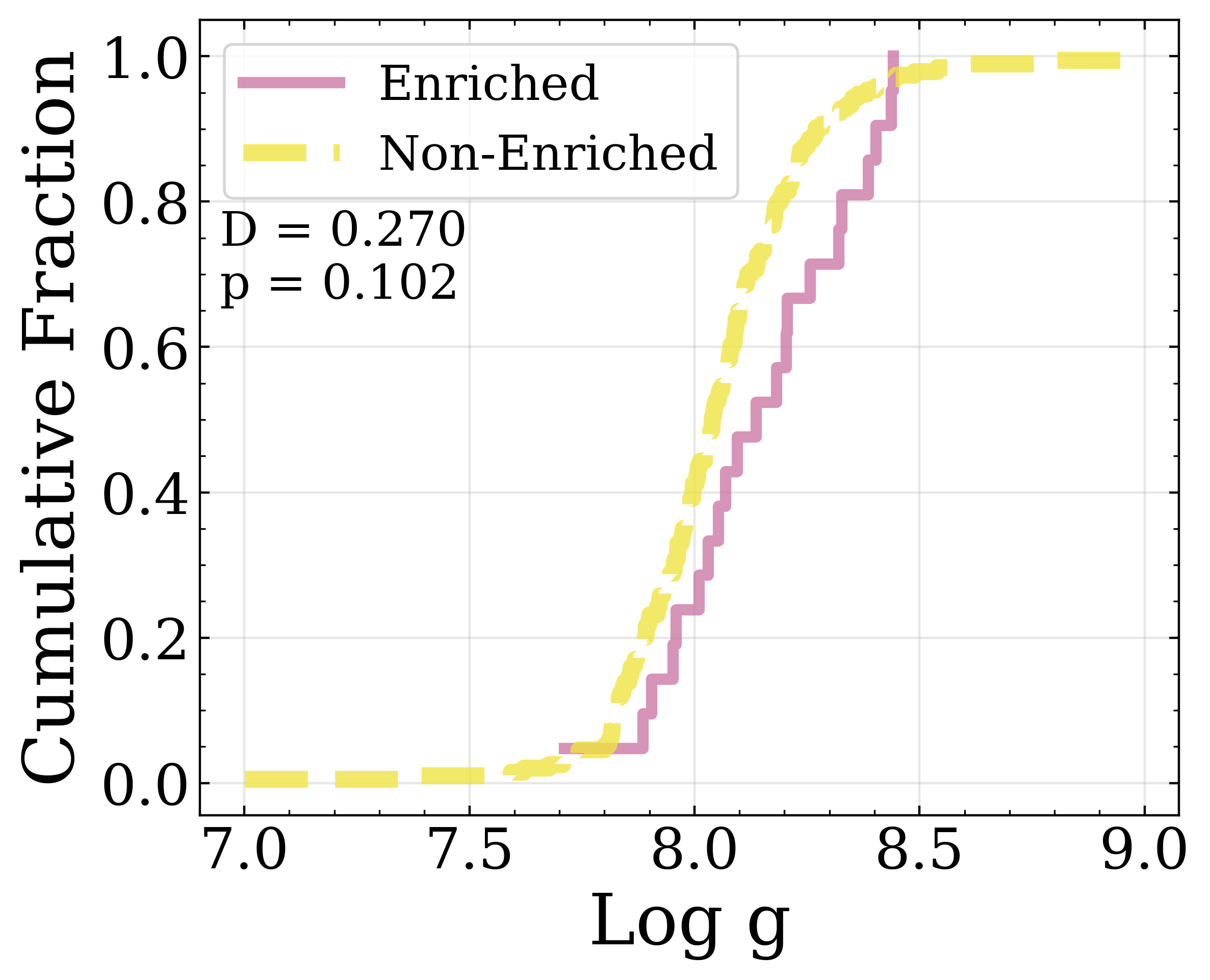}
\plottwo{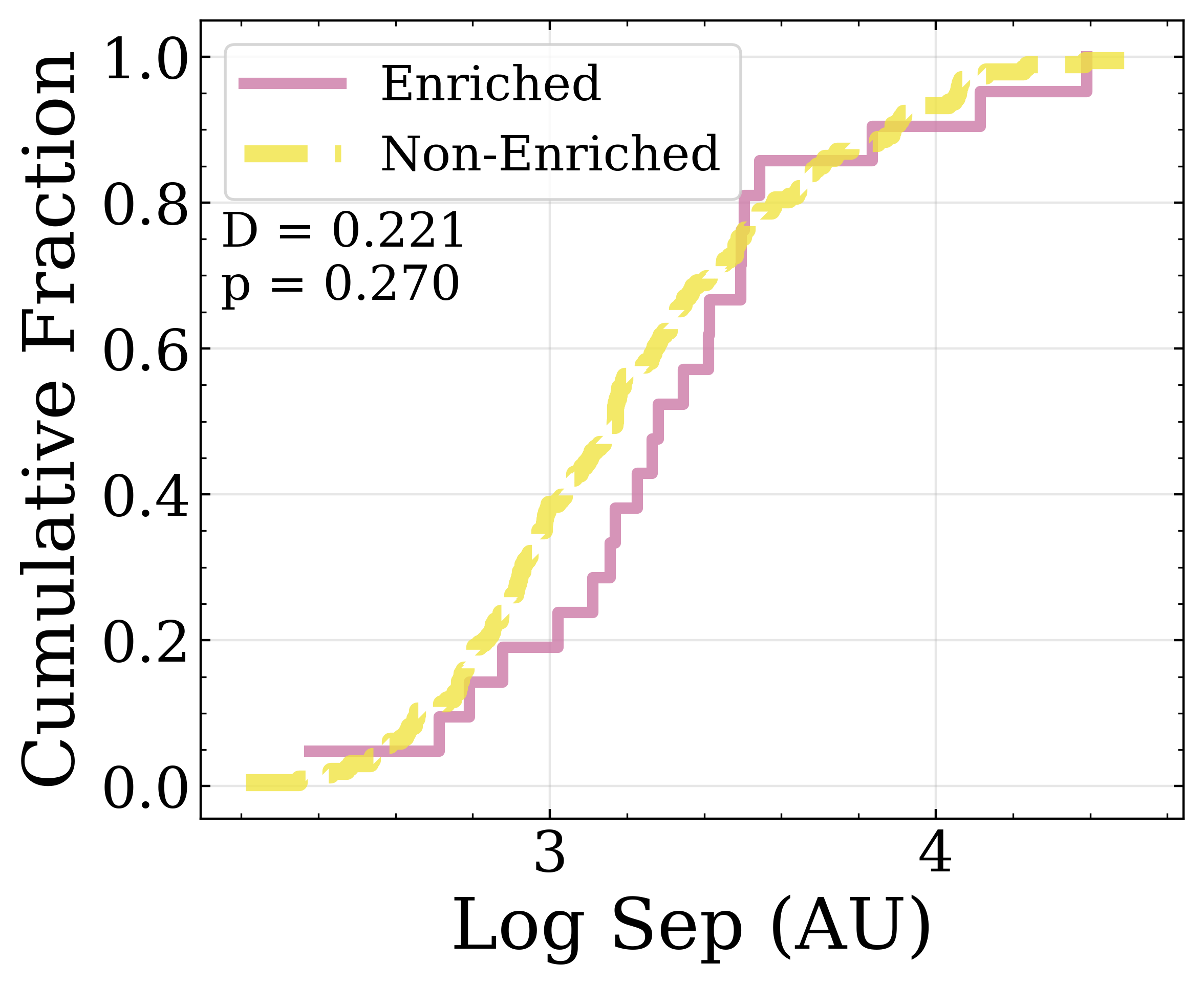}{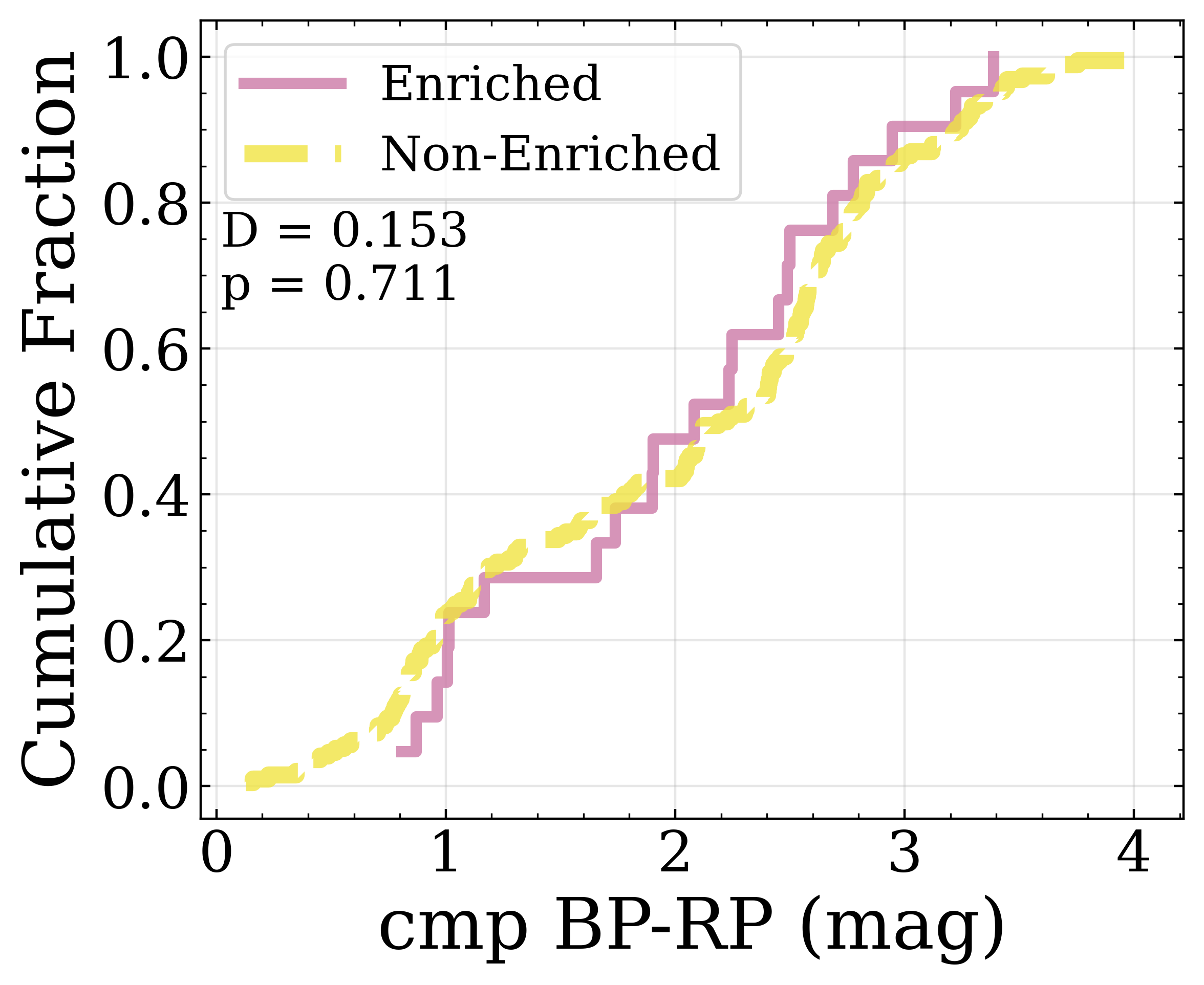}
\caption{Cumulative distributions of properties for metal-enriched (in pink) and non-enriched (in yellow) wide binary systems. {\color{hl} The result of the KS test is listed in each panel and the two samples are statistically indistinguishable.} The parameters include the \textit{Gaia} apparent G magnitude of the white dwarfs (D=0.323, p=0.029), the signal-to-noise ratio in the DESI blue arm (D=0.097, p=0.986), white dwarf temperature (D=0.209, p=0.331), surface gravity (D=0.257, p=0.137), projected separation (D=0.226, p=0.247), and the BP-RP color of the companions (D=0.143, p=0.776). 
}
\label{fig: binary properties}
\end{figure*}

We further investigate the properties of metal-enriched and non-enriched wide binary systems, as shown in Figure~\ref{fig: binary properties}. {\color{hl} A two-sample Kolmogorov-Smirnov (KS) test is performed to determine if the samples share the same underlying distribution. The resulting D-statistics and p-values are displayed in each panel, where a larger D-statistic and a smaller p-value indicate a higher probability that these two samples are drawn from different distributions.}  

In general, we find that the two samples are statistically similar. {\color{hl}We caution that the statistical power of the KS test is limited in detecting subtle differences with only 21 metal-enriched white dwarfs in wide binaries. For example, the dearth of bright enriched white dwarfs in wide binaries is likely a combination of small sample size and an artifact of the $\mathtt{fluxContamination\_DESI}$ quality cut described in Section~\ref{sec: sample}.} Brighter white dwarfs are more frequently associated with brighter companions, increasing the likelihood of contaminated DESI spectra and their subsequent removal from the sample. Moreover, the distribution of the signal-to-noise ratio in the blue arm ($\mathtt{SNmeasured\_B}$) is identical between the enriched and non-enriched samples, indicating that the detection of metal enrichment is not a result of higher-quality data or a fainter apparent magnitude. 

While the white dwarf temperature distributions appear distinct upon visual inspection, the KS test shows no statistically significant difference again, likely due to the small sample size. In Figure~\ref{fig: binary properties}, the distribution of surface gravity (log g) also shows a possible divergence, with metal-enriched systems having a higher surface gravity; however, this difference remains below the threshold for statistical significance. As discussed in \citet{Swan2026}, the white dwarf parameters (i.e., temperature and surface gravity) are derived using pure hydrogen or helium models, without considering metals. Therefore, these parameters, especially surface gravity, are less accurate for metal-enriched white dwarfs. We defer a detailed analysis of the metal enrichment dependence on white dwarf parameters to a future study. 

The projected binary separations extend from 160 to 33,640\,AU, and the metal-enriched and non-enriched systems share a similar distribution. Figure~\ref{fig: binary properties} also displays the distribution of \textit{Gaia} BP-RP colors for the companions of both enriched and non-enriched systems. This color serves as a proxy for the companions' effective temperatures, and their underlying distributions are found to be similar. This suggests that metal enrichment does not strongly depend on the physical properties of the companion star.

\begin{figure*}
\plottwo{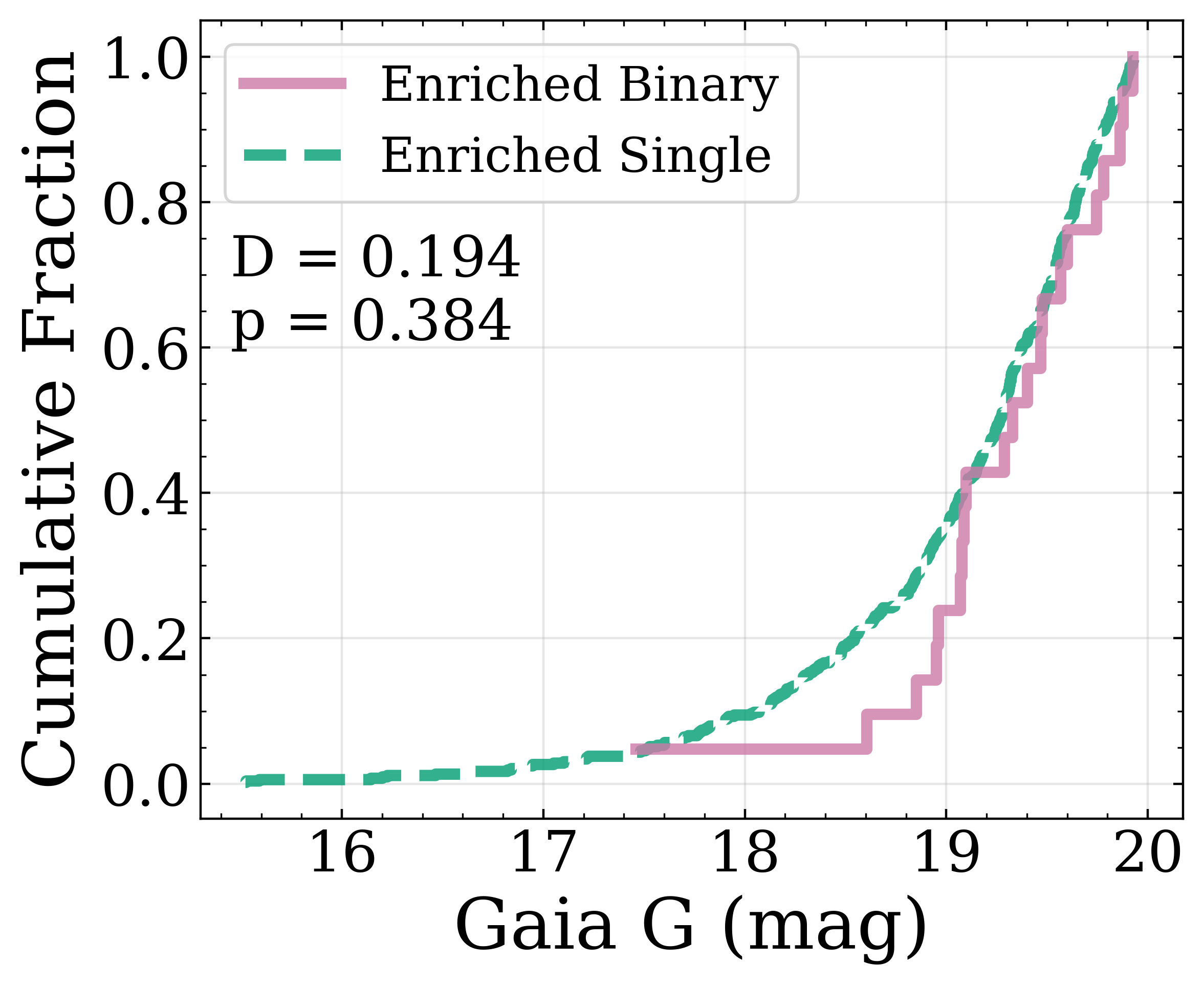}{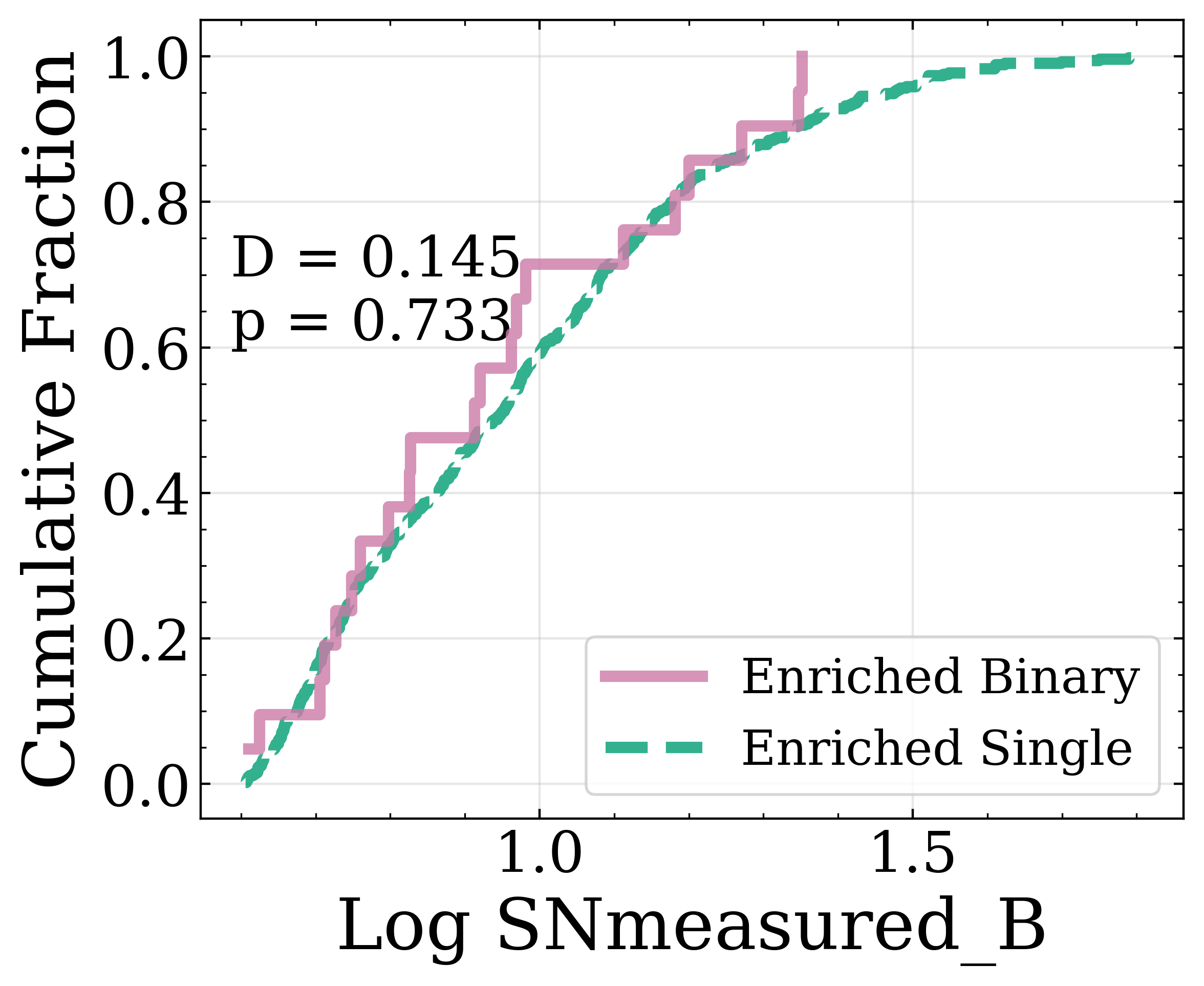}
\plottwo{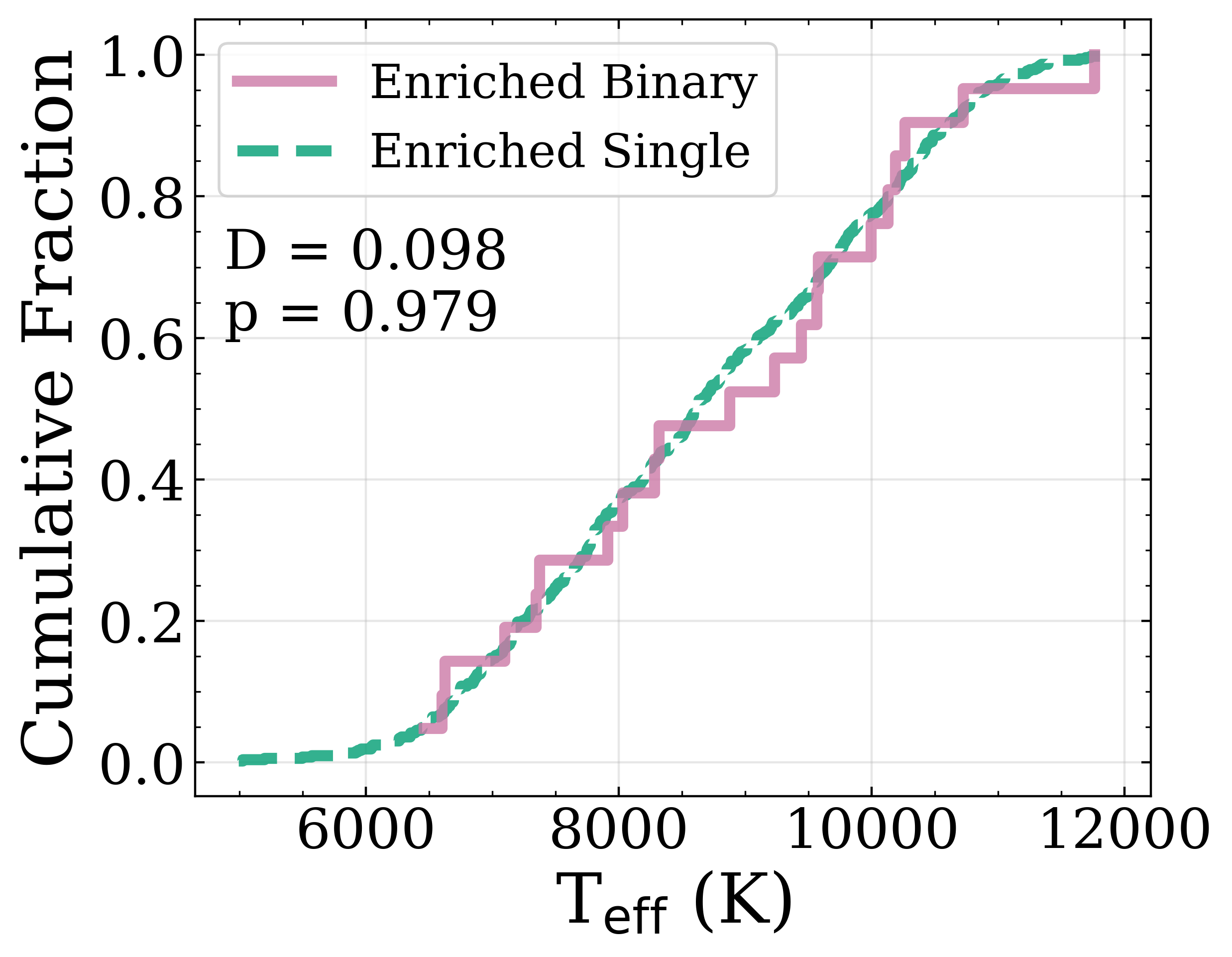}{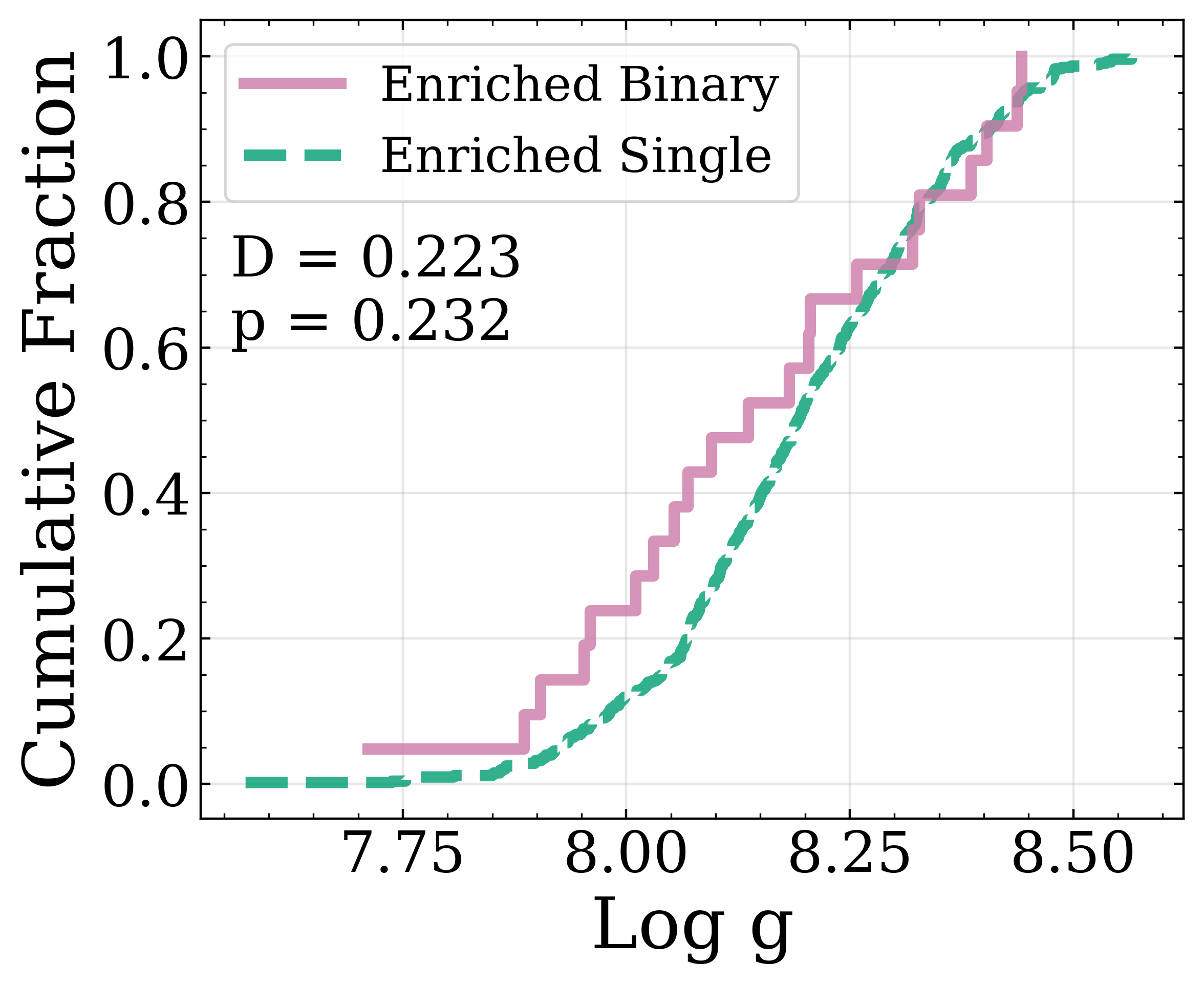}
\caption{
Cumulative distributions of metal-enriched white dwarfs in wide binaries (in pink) and single systems (in green). The KS test results are shown in each panel. The parameters include \textit{Gaia} apparent G magnitude (D=0.197, p=0.361), the signal-to-noise ration in the DESI blue arm (D=0.149, p=0.707), white dwarf temperature (D=0.094, p=0.987), and surface gravity (D=0.218, p=0.253). These two samples are statistically indistinguishable in these parameters. }
\label{fig: enriched properties}
\end{figure*}

Next we compare the properties of cool metal-enriched white dwarfs in wide binaries with those of single systems in Figure~\ref{fig: enriched properties}. {\color{hl} KS tests again show no statistically significant differences between the two samples across any of the considered parameters, suggesting that metal-enriched white dwarfs in wide binaries are similar to their single-star counterparts. Notably, the KS test shows that the surface gravity distributions of metal-enriched single white dwarfs and wide binaries are indistinguishable. This further supports the idea that the discrepancy in surface gravity seen in Figure~\ref{fig: binary properties} is likely a systematic effect arising from the white dwarf atmospheric models. Further investigation will be required once these parameters are more robustly determined, particularly for the enriched systems.}

\section{Discussion \label{sec: discussion}}

The large number of spectroscopic observations of white dwarfs in DESI DR1 \rf{--} combined with the high-quality, uniform data \rf{--} enabled the compilation of two robust comparison samples of cool, helium-dominated white dwarfs. Our main conclusion is that metal enrichment in wide binaries is significantly lower than that in single systems. 

\begin{figure}
\plotone{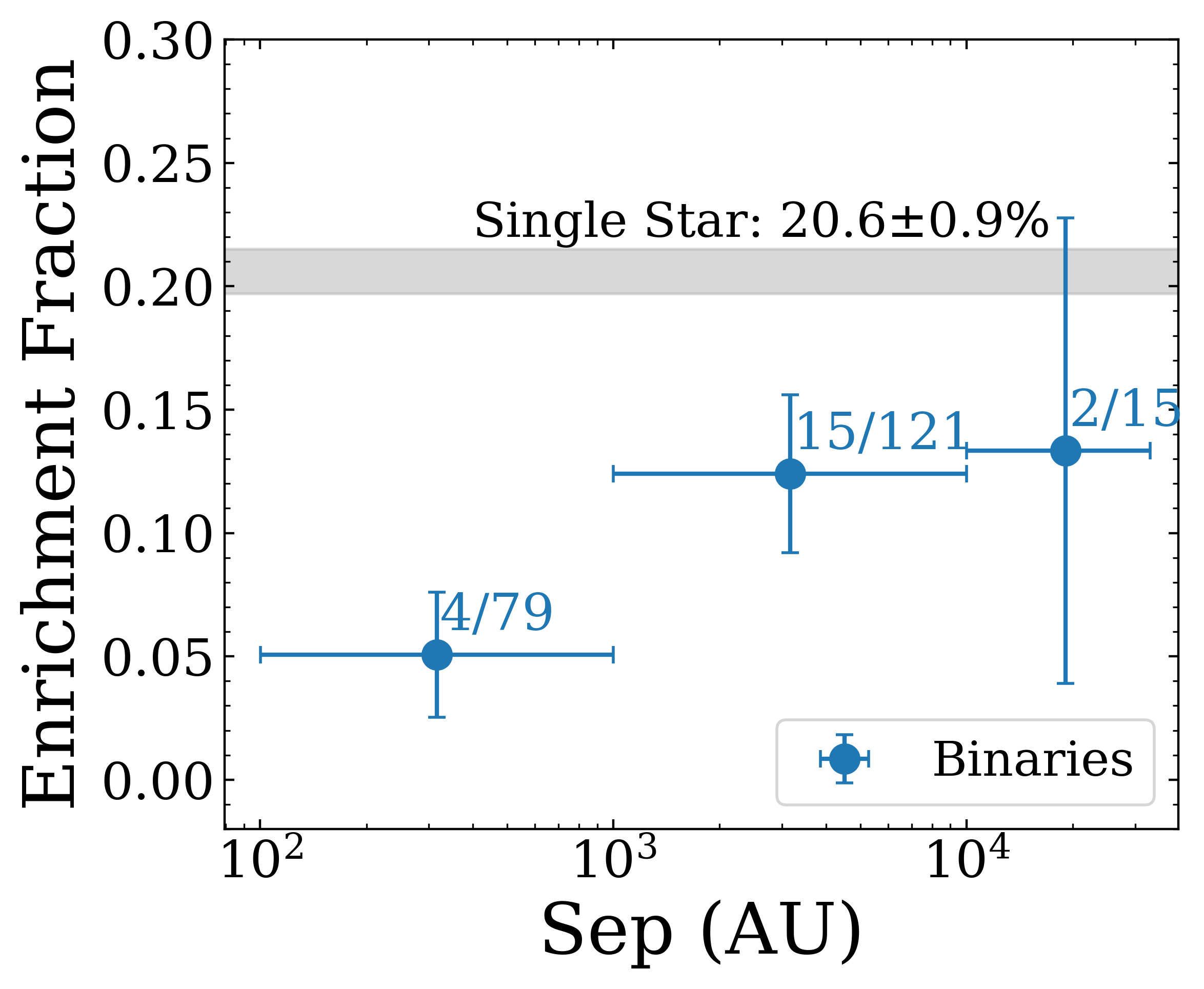}
\caption{Enrichment fraction in wide binaries as a function of projected separation. The fractions are 4/81 (100--1,000\,AU), 15/121 (1,000--10,000\,AU), and 2/15 ($>$ 10,000\,AU). The enrichment fraction of single white dwarfs is shown as the grey shaded area. The number of white dwarfs in each binary separation bin is listed in the figure. White dwarfs in wide binaries have a lower enrichment fraction compared to their single star counterparts. Furthermore, enrichment rate appears to be further suppressed around wide binaries with a projected separation less than $\approx$\,1000\,AU.
}
\label{fig: separation}
\end{figure}

We further investigate the enrichment fraction as a function of the projected separation. {\color{hl} As shown in Figure~\ref{fig: separation}, the enrichment rate exhibits a tentative dependence on separation. Specifically, the fraction increases from $4.9\,\pm\,2.5\%$ ($4/81$) for systems with projected separations below 1,000~AU to $12.4\,\pm\,3.2\%$ ($15/121$) for those between 1,000 and 10,000\,AU. While the difference is only at the $1.8\,\sigma$ level, our results are consistent with \citet{Zuckerman2014}, who also found a lower enrichment fraction ($1/21$; $5\%$) at smaller separations ($<2,500$\,AU) compared to wider systems ($5/17$; $30\%$ for $2,500\text{--}9,250$\,AU).} \citet{Williams2024} expanded on the wide binary sample from \citet{Zuckerman2014} and found that planetary accretion is further suppressed with a projected separation of 200/500\,AU during the main sequence/white dwarf stage; this result parallels the findings of the planet occurrence rate found around binary stars \citep{MoeKratter2021, Clark2024}. To calculate the metal enrichment fraction as a function of the binary separation on the main sequence is beyond the scope of the current work, given the issues with white dwarf parameters (see Section~\ref{sec: WD properties}); but it will be an interesting topic for future studies and to further connect metal-enriched white dwarfs to the planetary system architecture around main sequence stars. 

\citet{Noor2024} tackled the effect of binarity on metal enrichment from a different angle, and searched for wide binary companions across a range of orbital separations comparable to ours, utilizing several samples of known metal-enriched white dwarfs. For two DZ samples, including one covering effective temperature from 6,000 to 12,000\,K \citep{Dufour2007}, and the other from 4,500 to 9,000\,K \citep{Hollands2018}, they found that binary fraction is significantly smaller than that for field white dwarfs -- which is consistent with the findings in this paper.

On the other hand, \citet{Noor2024} found the binarity fraction for the warm DAZ sample (14,000\,$<$\,T$_\mathrm{eff}$\,$<$\,31,000\,K) appears to be indistinguishable from the field sample. \citet{Wilson2019} conducted a blind survey with \textit{Hubble} on hot white dwarfs (T$_\mathrm{eff}$\,$>$\,14,000~K), and found that 8 out of 12 wide binary targets show atmospheric enrichment, resulting in an enrichment fraction of 67$^{+10}_{-15}$\%, broadly consistent with a rate of 40--50\% around single white dwarfs over a similar temperature range observed with \textit{Hubble}\footnote{\textit{Hubble} observations are significantly more sensitive to metal enrichment than ground-based observations; therefore, the reported fractions from these studies are typically higher than the 20\% quoted in Section~\ref{sec: introduction}.} \citep{OuldRouis2024, Koester2014a}.

The emerging picture is that in warm white dwarfs (T$_\mathrm{eff}$\,$\gtrsim$\,14,000\,K, {\color{hl} therefore young white dwarfs with shorter cooling ages}), the fraction of metal enrichment in wide binaries \rf{appears to be} indistinguishable from that of single systems. In contrast, for cooler systems (T$_\mathrm{eff}$\,$\lesssim$\,11,500\,K, {\color{hl} therefore old white dwarfs with longer cooling ages}), the enrichment fraction in wide binaries is much lower than that in single systems. \rf{At least two potential explanations exist for the dearth of unstable planetesimals around cool white dwarfs in binaries.} 

\rf{\textit{One possibility is that binary systems have a smaller initial reservoir of planetary material compared to single stars.} The lack of remaining unstable planetesimals results in the lower enrichment levels seen in the cooler wide-binary white dwarfs studied here. This ``small reservoir" hypothesis is further supported by the findings that planet and planetesimal formation appears to be inhibited in binaries, particularly those with separations $\lesssim$\,200\,AU \citep[e.g.,][]{Yelverton2019, MoeKratter2021}. }


\rf{\textit{Another possibility is that binary systems start out with planetary material reservoirs similar to those of single stars but deplete them more rapidly during the white dwarf stage.} Most of the literature suggests that binarity leads to an increase in tidal disruption events \citep[e.g.,][]{Bonsor2015, Stephan2017, Trierweiler2026}, \rf{in contrast to our findings in cool systems}. It is possible that tidal disruption events in wide binaries occur much earlier in the evolution of the white dwarf; such an increase in planetary accretion might therefore be more evident in a sample of hotter white dwarfs with even shorter cooling ages than the warm sample ($\gtrsim$\,14,000\,K) discussed here. A higher rate of tidal disruption events earlier on can deplete unstable planetesimals and 
potentially produce the lower enrichment fraction seen in cooler systems.}


\rf{These two mechanisms may operate in combination. A key observational test is to move beyond simple metal enrichment fraction calculations to properly model each system and determine its level of metal enrichment, similar to \citealt{Lizana-Vidal2026}. Comparing the magnitude and age dependence of this enrichment between single and binary systems will distinguish between these two scenarios. }

Another application of this wide binary sample is to test the hypothesis that planets share the same chemical fingerprints as \rf{their} host star \citep{Thiabaud2015}. Such measurements are difficult to do directly on exoplanets due to the strong degeneracies on planet composition \citep{Searger2007, Dorn2015}, but it is possible using metal-enriched white dwarfs in wide binaries. Wide binaries are presumably chemically homogeneous, therefore they provide a direct test by comparing the planetary composition from metal-enriched white dwarfs with the stellar composition directly from the main sequence star. Pilot work has been done on a few systems, and it seems that planetary composition follows that of its host star \citep{Bonsor2021, AguileraGomez2025}. On the other hand, \citet{Jenkins2024} did not find any difference in the metallicity of the companion star between the metal-enriched and non-enriched white dwarfs, \rf{suggesting that gas giants might only play a minor role in white dwarf accretion events}. {\color{hl} The wide binary sample in the Appendix A is a prime sample to further investigate this effect}. 

\section{Conclusions \label{sec: conclusions}}

In this study, \rf{two control samples} of cool helium-dominated white dwarfs (temperature from 5,000\,K to 11,500\,K, cooling age between 460\,Myr and 6.4\,Gyr) \rf{indicate} that the fraction of metal enrichment is significantly lower in wide binaries compared to that in single stars. \rf{The lower metal enrichment fraction is also recovered in the comparison SDSS sample, albeit at lower statistical significance.} In addition, there appears to be a dependence of metal enrichment on projected separation, with the metal enrichment further suppressed in binaries with a separation $\lesssim$ 1000\,AU. 
The implication is that \rf{compared to single stars, binary systems either possess a smaller initial reservoir of planetary material or deplete that material more rapidly. A key observational test to distinguish between these two scenarios is to probe the level of metal enrichment and its age dependence across single and binary systems.}

\rf{Indeed, previous studies have found that the binary fraction of warm ($T_\mathrm{eff} \gtrsim 14,000\,\text{K}$), metal-enriched white dwarfs is comparable to that of field white dwarfs. It would be valuable to directly measure the enrichment fraction and accretion rates in an unbiased sample of warm white binaries in wide binaries to test whether stellar companions drive more rapid depletion of planetary material during the white dwarf phase.}
\rf{More generally, our} results suggest that stellar binarity plays an important role in shaping the long-term evolution and survival of planetary systems, a possibility that future observations and dynamical studies will help to clarify.

\begin{acknowledgments}

The authors thank the anonymous referee for helpful comments that improved the quality of the manuscript. S. Xu thanks Eric W. Peng and Ben Zuckerman for insightful conversations and discussions during the preparation of the manuscript. Google Gemini and ChatGPT were used to improve the clarity and grammar of the manuscript. 

S. Xu \rf{is} supported by the international Gemini Observatory, a program of NSF NOIRLab, which is managed by the Association of Universities for Research in Astronomy (AURA) under a cooperative agreement with the U.S. National Science Foundation, on behalf of the Gemini partnership of Argentina, Brazil, Canada, Chile, the Republic of Korea, and the United States of America. L. K. Rogers and J. Najita are supported by NOIRLab, which is managed by the Association of Universities for Research in Astronomy (AURA) under a cooperative agreement with the U.S. National Science Foundation. 

This work is also partly supported by HST \# 17185, which was provided by NASA through a grant from the Space Telescope Science Institute, which is operated by the Association of Universities for Research in Astronomy, Inc., under NASA contract NAS 5-26555.

This research received funding from the European Research Council under the European Union’s Horizon 2020 research and innovation programme numbers 101020057 (BTG, PI, AS).

This material is based upon work supported by the U.S. Department of Energy (DOE), Office of Science, Office of High-Energy Physics, under Contract No. DE–AC02–05CH11231, and by the National Energy Research Scientific Computing Center, a DOE Office of Science User Facility under the same contract. Additional support for DESI was provided by the U.S. National Science Foundation (NSF), Division of Astronomical Sciences under Contract No. AST-0950945 to the NSF’s National Optical-Infrared Astronomy Research Laboratory; the Science and Technology Facilities Council of the United Kingdom; the Gordon and Betty Moore Foundation; the Heising-Simons Foundation; the French Alternative Energies and Atomic Energy Commission (CEA); the National Council of Humanities, Science and Technology of Mexico (CONAHCYT); the Ministry of Science, Innovation and Universities of Spain (MICIU/AEI/10.13039/501100011033), and by the DESI Member Institutions: \url{https://www.desi.lbl.gov/collaborating-institutions}. Any opinions, findings, and conclusions or recommendations expressed in this material are those of the author(s) and do not necessarily reflect the views of the U. S. National Science Foundation, the U. S. Department of Energy, or any of the listed funding agencies.

The authors are honored to be permitted to conduct scientific research on I'oligam Du'ag (Kitt Peak), a mountain with particular significance to the Tohono O'odham Nation.

\end{acknowledgments}

\facilities{DESI}

\software{numpy \citep{Numpy},
Astropy \citep{astropy:2013, astropy:2018, astropy:2022},
matplotlib \citep{Matplotlib},
astroquery \citep{Astroquery},
scipy \citep{Scipy}
}

\begin{deluxetable*}{lcclclcc}
\tabletypesize{\scriptsize}
\tablewidth{0pt} 
\label{tab: WDMS}
\tablecaption{This table lists properties of the 21 metal-enriched DZs and their wide binary companions (cmp). The white dwarf parameters, including its name $\mathrm{WDJname}$, temperature $\mathrm{Teff}$, and surface gravity $\mathrm{log g}$ are all from the DESI DR1 white dwarf catalog \citep{Swan2026}. The other parameters are from the \textit{Gaia} DR3 binary catalog \citep{El-Badry2021}. \label{tab: binary}}
\tablehead{
\colhead{WDJname} & \colhead{Teff (K)}& \colhead{logg} & \colhead{WD Gaia ID} & \colhead{WD G} & \colhead{cmp Gaia ID} & \colhead{cmp G} & \colhead{Binary Type}
}
\startdata 
WDJ001303.36+030304.88 & 7099 & 8.3 & Gaia DR3 2740492325179983872 & 19.1 & Gaia DR3 2740492329476127360 & 17.9 & WDMS \\
WDJ002433.70+034907.00 & 10724 & 8.1 & Gaia DR3 2548740039916337536 & 19.0 & Gaia DR3 2548740044213147904 & 17.9 & WDMS \\
WDJ003244.21+082720.22 & 6604 & 8.1 & Gaia DR3 2749781068130558464 & 19.5 & Gaia DR3 2749781072426150784 & 13.1 & WDMS \\
WDJ003917.52+232753.33 & 11763 & 8.2 & Gaia DR3 2806310603047483904 & 19.0 & Gaia DR3 2806310641702070016 & 16.3 & WDMS \\
WDJ012454.23-033519.83 & 7914 & 7.9 & Gaia DR3 2484109681283815168 & 19.1 & Gaia DR3 2484109784363031296 & 19.3 & WDMS \\
WDJ020211.42+065608.90 & 8033 & 8.3 & Gaia DR3 2567791793287585536 & 19.6 & Gaia DR3 2567791797581989120 & 16.6 & WDMS \\
WDJ075421.04+495016.68 & 10189 & 8.2 & Gaia DR3 934175968363650048 & 19.5 & Gaia DR3 934175968363647360 & 20.4 & WDWD \\
WDJ080131.13+532900.68 & 8282 & 8.0 & Gaia DR3 936641833642799744 & 17.5 & Gaia DR3 936641902362276736 & 13.1 & WDMS \\
WDJ101600.26+270154.63 & 9445 & 8.0 & Gaia DR3 739321312258441216 & 19.7 & Gaia DR3 739321307963291136 & 17.6 & WDMS \\
WDJ102625.84-050243.56 & 6459 & 8.4 & Gaia DR3 3779248273299114112 & 19.8 & Gaia DR3 3779248277594701952 & 13.1 & WDMS \\
WDJ120428.59+503317.18 & 10129 & 8.2 & Gaia DR3 1546908475396468992 & 18.9 & Gaia DR3 1546908960731418240 & 15.7 & WDMS \\
WDJ122102.43+554835.08 & 9996 & 8.1 & Gaia DR3 1574343386453534208 & 19.3 & Gaia DR3 1574343390748734592 & 13.4 & WDMS \\
WDJ123436.77-021645.25 & 6624 & 7.7 & Gaia DR3 3683196449325908480 & 18.6 & Gaia DR3 3683196453620955904 & 11.0 & WDMS \\
WDJ123711.25+545128.03 & 8320 & 8.0 & Gaia DR3 1571089381790648064 & 19.6 & Gaia DR3 1571089386086876032* & 16.1 & WDMS \\
WDJ131529.25+522810.82 & 9234 & 8.3 & Gaia DR3 1563055117184297984 & 19.9 & Gaia DR3 1563055117183845120 & 16.9 & WDMS \\
WDJ170644.59+270024.67 & 9568 & 8.1 & Gaia DR3 4574839773168096640 & 19.9 & Gaia DR3 4574839807529446784 & 13.3 & WDMS \\
WDJ174213.13+155719.84 & 7375 & 8.4 & Gaia DR3 4549217677612386048 & 19.1 & Gaia DR3 4549217681907301504 & 20.5 & WDWD \\
WDJ203434.52-031523.42 & 8880 & 7.9 & Gaia DR3 4224790141724412160 & 19.9 & Gaia DR3 4224790141723498368 & 12.5 & WDMS \\
WDJ214536.09+071204.78 & 9579 & 8.4 & Gaia DR3 2700845756950789504 & 19.4 & Gaia DR3 2700847234417809792 & 15.5 & WDMS \\
WDJ222207.51+220643.14 & 7346 & 8.4 & Gaia DR3 1874965499168713600 & 19.3 & Gaia DR3 1874965739686844032 & 13.3 & WDMS \\
WDJ235544.80+294619.34 & 10264 & 8.0 & Gaia DR3 2867316975650195968 & 19.1 & Gaia DR3 2867316941290457856 & 14.7 & WDMS \\
\enddata
\tablecomments{$^*$ There is \rf{a} DESI spectrum available for this object in the DR1 stellar catalog \citep{Koposov2024}.}
\end{deluxetable*}

{\bf Data Availability}

The DESI DR1 data presented here all available from the public archives. Data used to produce the figures presented in this paper are available under DOI: 10.5281/zenodo.22133520 and can be found here:  \href{https://zenodo.org/records/22133520}{https://zenodo.org/records/22133520}.

\end{CJK}
\bibliography{WD_Master.bib}{}
\bibliographystyle{aasjournalv7}



\end{document}